\documentclass[aip,jcp,twocolumn,showpacs,amsmath,amssymb,floatfix,reprint]{revtex4-1}

\usepackage{graphicx}
\usepackage[bookmarks,bookmarksopen,bookmarksnumbered,colorlinks,linkcolor=red,linktocpage,citecolor=blue,urlcolor=cyan,pdfpagemode=UseOutline]{hyperref}
\usepackage{color}
\usepackage{epstopdf}

\def\bea{\begin{eqnarray}}
\def\eea{\end{eqnarray}}
\def\ben{\begin{equation}}
\def\een{\end{equation}}
\def\benu{\begin{enumerate}}
\def\enu{\end{enumerate}}

\def\bei{\begin{itemize}}
\def\eei{\end{itemize}}
\def\benu{\begin{enumerate}}
\def\enu{\end{enumerate}}

\def\n{n}

\def\sss{\scriptscriptstyle\rm}

\def\1var{(\bx_1...\bx\N)}

\def\br{{\bf r}}

\def\bx{{x}}

\def\x{_{\sss X}}
\def\c{_{\sss C}}
\def\s{_{\sss S}}
\def\xc{_{\sss XC}}
\def\Hx{_{\sss HX}}

\def\N{_{\sss N}}
\def\H{_{\sss H}}

\def\ee{_{\rm ee}}

\def\sph_int{ {\int d^3 r}}

\newcommand{\intd}{\mathrm{d}}

\newcommand{\matelem}[3]{\left\langle #1 \left| #2 \right| #3 \right\rangle}

\providecommand{\abs}[1]{\left|#1\right|}

\newcommand{\parref}[1]{(\ref{#1})}

\DeclareMathOperator{\tr}{tr}

\providecommand{\bra}[1]{\left< #1 \right|}
\providecommand{\ket}[1]{\left| #1 \right>}

\def\wt{\texttt{w}}
\def\wtb{\overline{\wt}}

\def\Hxw{_{{\sss HX},\wt}}
\def\xcw{_{{\sss XC},\wt}}
\def\xw{_{{\sss X},\wt}}
\def\Hw{_{{\sss H},\wt}}

\def\wtset{{\sss\mathcal{W}}}
\def\Hxcwset{_{{\sss HXC},\wtset}}
\def\Hxwset{_{{\sss HX},\wtset}}
\def\xcwset{_{{\sss XC},\wtset}}

\def\cwset{_{{\sss C},\wtset}}
\def\swset{_{{\sss S},\wtset}}

\begin{document}

\title{Excitations and benchmark ensemble density functional theory for two electrons}
\author{Aurora Pribram-Jones}
\affiliation{Department of Chemistry, University of
California-Irvine, Irvine, CA 92697, USA}
\author{Zeng-hui Yang}
\affiliation{Department of Physics and Astronomy, University of
Missouri, Columbia, MO 65211, USA}
\author{John R. Trail}
\affiliation{Theory of Condensed Matter Group, Cavendish Laboratory, University of Cambridge, Cambridge CB3 0HE, United Kingdom}
\author{Kieron Burke}
\affiliation{Department of Chemistry, University of
California-Irvine, Irvine, CA 92697, USA}
\author{Richard J. Needs}
\affiliation{Theory of Condensed Matter Group, Cavendish Laboratory, University of Cambridge, Cambridge CB3 0HE, United Kingdom}
\author{Carsten A. Ullrich}
\affiliation{Department of Physics and Astronomy, University of
Missouri, Columbia, MO 65211, USA}
\date{\today}

\begin{abstract}
A new method for extracting ensemble Kohn-Sham potentials from 
accurate excited state densities is applied to a variety of two
electron systems, exploring the behavior of exact ensemble
density functional theory.   The issue of separating the Hartree
energy and the choice of degenerate eigenstates is explored.
A new approximation, spin eigenstate Hartree-exchange (SEHX), is
derived.  Exact conditions that are proven include the signs
of the correlation energy components, the virial theorem for
both exchange and correlation, and the asymptotic behavior of
the potential for small weights of the excited states.
Many energy components are given as a function of the weights
for two electrons in a one-dimensional flat box, in a box
with a large barrier to create charge transfer excitations,
in a three-dimensional harmonic well (Hooke's atom), and
for the He atom singlet-triplet ensemble, singlet-triplet-singlet ensemble, and triplet bi-ensemble.  
\end{abstract}

\maketitle
%\tableofcontents

\section{Introduction and illustration}
Ground-state density functional theory\cite{HK64,KS65} (DFT) is a popular
choice for finding the ground-state energy of electronic
systems,\cite{B12} and excitations can now easily be extracted using
time-dependent DFT\cite{RG84,C96,MMNG12,U12} (TDDFT).  Despite its popularity, TDDFT calculations have many
well-known failings,\cite{ORR02,MZCB04,HIRC11,UY13} such as double excitations\cite{EGCM11} and charge-transfer excitations.\cite{DWH03,T03} Alternative DFT treatments of
excitations\cite{G96,FHMP98,LN99} are always of interest.

Ensemble DFT (EDFT)\cite{T79,GOKb88,GOK88,OGKb88} is one such alternative
approach. Unlike TDDFT, it is based on
an energy variational principle.  An ensemble of monotonically
decreasing weights is constructed from the $M+1$
lowest levels of the system, and the expectation value
of the Hamiltonian over orthogonal trial wavefunctions is
minimized by the $M+1$ exact lowest eigenfunctions.\cite{GOKb88}
A one-to-one correspondence can be established between ensemble
densities and potentials for a given set of weights, providing
a Hohenberg-Kohn theorem, and application to non-interacting
electrons of the same ensemble density yields a Kohn-Sham scheme
with corresponding equations.\cite{GOK88}  In principle, this yields the
exact ensemble energy, from which individual excitations may be
extracted.

But to make a practical scheme, approximations must be used.\cite{N95,N98,SD99,N01,FF13}
These have been less successful for EDFT than those of ground-state\cite{B88,LYP88,B93,PBE96,PBE98} and TDDFT,\cite{JWPA09,MMNG12} and their accuracy is not yet competitive with TDDFT transition frequencies from standard approximations.   Some progress has been
made in identifying some major sources of error.\cite{GPG02,TN03,TNb03}

To help speed up that progress, we have developed a numerical
algorithm to calculate ensemble Kohn-Sham (KS) quantities (orbital energies,
energy components, potentials, etc.) essentially exactly,\cite{YTPB14} from
highly accurate excited-state densities.  In the present paper,
we provide reference KS calculations and results
for two-electron systems under a variety of conditions.  The potentials we find differ
in significant ways from the approximations suggested so far,
hopefully leading to new and better approximations.

\begin{figure}[htbp]
\includegraphics[height=0.85\columnwidth,angle=-90]{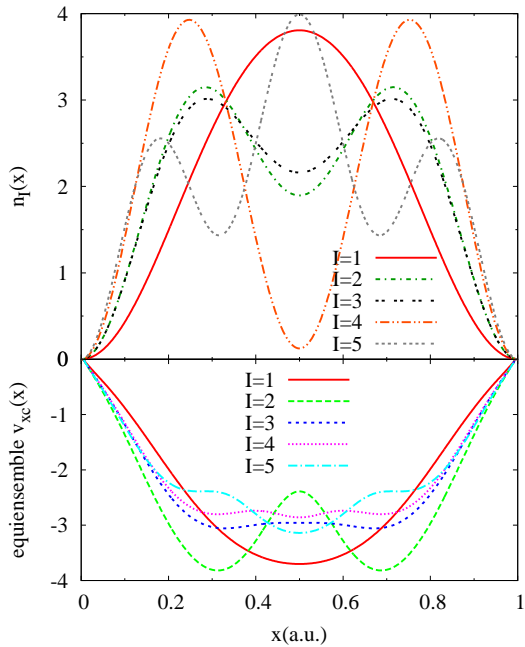}
\caption{Exact densities and equiensemble exchange-correlation potentials of the 1D box with two electrons. The third excited state ($I=4$) is a double excitation.  See Sec. \ref{sect:1dbox}.}
\label{fig:1dbox:general}
\end{figure}

To illustrate the essential idea, we perform calculations
on simple model systems.  For example, Sec. \ref{sect:1dbox} presents two `electrons' in a one-dimensional box,
repelling one another via a (slightly softened) Coulomb repulsion.
In Fig. \ref{fig:1dbox:general}, we show their ground- and excited-state densities,
with $I$ indicating the specific ground or excited state.  We also plot the
ensemble exchange-correlation potentials for equally weighted mixtures of the ground and excited states, which result from our inversion scheme.  In this lower plot, $I=1$ denotes the ground-state exchange-correlation potential, and $I>1$ indicates the potential corresponding to an equal mixture of the ground state and all multiplets up to and including the $I$-th state.  Excitation energies for all these states are extracted using the EDFT methods described below.

The paper is laid out as follows.  In the next section, we briefly
review the state-of-the-art for EDFT, introducing our notation.
Then we give some formal considerations about how to define the
Hartree energy.  The naive definition, taken directly from ground-state
DFT, introduces spurious unphysical contributions (which then must be
corrected-for) called `ghost' corrections.\cite{GPG02}  We also consider
how to make choices among KS eigenstates when they are degenerate,
and show that such choices matter to the accuracy of the approximations.
We close that section by showing how to construct symmetry-projected
ensembles.

In the following section, we prove a variety of exact conditions
within EDFT.  Such conditions have been vital in constructing
useful approximations in ground-state DFT.\cite{LP85,PBE96}  Following that, we describe
our numerical methods in some detail.

The results section consists of calculations for quite distinct
systems, but all with just two electrons.  The one-dimensional
flat box was used for the illustration here, which also gives rise
to double excitations.  A box with a high, asymmetric barrier produces charge-transfer excitations. Hooke's atom is a three-dimensional
system, containing two Coulomb-repelling electrons in a
harmonic oscillator external potential.\cite{FUT94}  It has proven useful in
the past to test ideas and approximations in both ground-state
and TDDFT calculations.\cite{HMB02}  We close the section reporting several
new results for the He atom, using ensembles that include low-lying triplet states.
Atomic units [$e=\hbar=m_e=1/(4\pi\epsilon_0)=1$] are used throughout unless
otherwise specified.

\section{Background}
\label{sect:theory}
\subsection{Basic theory}
The ensemble variational principle\cite{GOKb88} states that, for an ensemble of the lowest $M+1$ eigenstates $\Psi_0,\ldots,\Psi_M$ of the Hamiltonian $\hat{H}$ and a set of orthonormal trial functions $\tilde{\Psi}_0,\ldots,\tilde{\Psi}_M$,
\ben
\sum_{m=0}^M \wt_m\matelem{\tilde{\Psi}_m}{\hat{H}}{\tilde{\Psi}_m}\ge \sum_{m=0}^M \wt_mE_m,
\een
when the set of weights $\wt_m$ satisfies
\ben
\wt_0\ge \wt_1\ge\ldots\ge\wt_m\ge\ldots\ge0,
\een
and $E_m$ is the eigenvalue of the $m$th eigenstate of $\hat{H}$. Equality holds only for $\tilde{\Psi}_m=\Psi_m$. The density matrix of such an ensemble is defined by
\ben
\hat{D}_{\wtset}=\sum_{m=0}^M \wt_m\ket{\Psi_m}\bra{\Psi_m},
\een
where $\wtset$ denotes the entire set of weight parameters. Properties of the ensemble are then defined as traces of the corresponding operators with the density matrix. The ensemble density $n_\wtset(\br)$ is
\ben
n_{\wtset}(\br)=\tr\{\hat{D}_{\wtset}\hat{n}(\br)\}=\sum_{m=0}^M \wt_m n_m(\br),
\label{eqn:theory:den}
\een
and the ensemble energy $E_\wtset$ is
\ben
E_{\wtset}=\tr\{\hat{D}_{\wtset}\hat{H}\}=\sum_{m=0}^M \wt_m E_m.
\label{eqn:theory:EensDef}
\een
$n_\wtset(\br)$ is normalized to the number of electrons, implying $\sum_{m=0}^M \wt_m=1$.

A Hohenberg-Kohn (HK)\cite{HK64} type theorem for the one-to-one correspondence between $n_\wtset(\br)$ and the potential in $\hat{H}$ has been proven,\cite{T79,GOK88} so all ensemble properties are functionals of $n_\wtset(\br)$, including $\hat{D}_{\wtset}$. The ensemble HK theorem allows the definition of a non-interacting KS system, which reproduces the exact $n_\wtset(\br)$. The existence of an ensemble KS system assumes ensemble $v$-representability. EDFT itself, however, only requires ensemble non-interacting $N$-representability, since a constrained-search formalism is available.\cite{GOK88} Ensemble $N$- and $v$-representability are not yet proven, only assumed.

As in the ground-state case, only the ensemble energy functional is formally known, which is
\ben
E_{\wtset}[n]=F_{\wtset}[n]+\int\intd^3r\;n(\br)v(\br),
\een
where $v(\br)$ is the external potential. The ensemble universal functional $F_\wtset$ is defined as
\ben
F_{\wtset}[n]=\tr\{\hat{D}_{\wtset}[n](\hat{T}+\hat{V}\ee)\},
\een
where $\hat{T}$ and $\hat{V}\ee$ are the kinetic and electron-electron interaction potential operators,
respectively. The ensemble variational principle ensures that the ensemble energy functional evaluated at the exact ensemble density associated with $v(\br)$ is the minimum of this functional, Eq. \parref{eqn:theory:EensDef}.

The ensemble KS system is defined as the non-interacting system that reproduces $n_\wtset(\br)$ and satisfies the following non-interacting Schr\"{o}dinger equation:
\ben
\left\{-\frac{1}{2}\nabla^2+v_{\text{s},\wtset}[n_{\wtset}](\br)\right\}\phi_{j,\wtset}(\br)=\epsilon_{j,\wtset}\phi_{j,\wtset}(\br).
\label{eqn:theory:KS}
\een
The ensemble KS system has the same set of $\wt_m$ as the interacting system. There is no formal proof for this consistency, and it has non-trivial implications even for simple systems. This will be explored more in Sec. \ref{sect:theory:connection}.

The KS density matrix $\hat{D}_{\text{s},\wtset}$ is
\ben
\hat{D}_{\text{s},\wtset}=\sum_{m=0}^M \wt_m\ket{\Phi_m}\bra{\Phi_m},
\een
where $\Phi_m$ are non-interacting $N$-particle wavefunctions, usually assumed to be single Slater determinants formed by KS orbitals $\phi_{j,\wtset}$. We find that this choice can be problematic, and it will be discussed in Sec. \ref{sect:theory:Hartree}. The ensemble density $n_\wt(\br)$ is reproduced by the KS system, meaning
\ben
n_{\wtset}(\br)=\sum_{m=0}^M \wt_m n_m(\br)=\sum_{m=0}^M \wt_m n_{\text{s},m}(\br),
\label{eqn:theory:nens}
\een
where $n_m(\br)=\matelem{\Psi_m}{\hat{n}(\br)}{\Psi_m}$, and $n_{\text{s},m}(\br)=\matelem{\Phi_m}{\hat{n}(\br)}{\Phi_m}$. The KS densities of the individual states are generally not related to those of the interacting system; only their weighted sums are equal, as in Eq. \parref{eqn:theory:nens}.

$E_{\wtset}[n]$ is decomposed as in ground-state DFT,
\ben
\begin{split}
E_{\wtset}[n]&=T\swset[n]+V[n]+E\H[n]+E\xcwset[n]\\
&=\tr\{\hat{D}\swset\hat{T}\}+\int\intd^3r\;n(\br)v(\br)\\
&\quad+E\H[n]+E\xcwset[n],
\end{split}
\label{eqn:theory:Edecomp}
\een
where only the ensemble exchange-correlation (XC) energy $E\xcwset$ is unknown. The form of $v\swset(\br)$ is then determined according to the variational principle by requiring $\delta E_{\wtset}[n_{\wtset}]/\delta n_{\wtset}(\br)=0$, resulting in
\ben
v\swset[n_{\wtset}](\br)=v(\br)+v\H[n_{\wtset}](\br)+v\xcwset[n_{\wtset}](\br),
\een
where $v\H[n](\br)=\delta E\H[n]/\delta n(\br)$, and $v\xcwset[n](\br)=\delta E\xcwset[n]/\delta n(\br)$. $E\H$ is generally defined to have the same form as the ground-state Hartree energy functional. Although this choice is reasonable, we find that it is more consistent to consider $E\Hx$, the combined Hartree and exchange energy. This point will be discussed in Sec. \ref{sect:theory:Hartree}.

The ensemble universal functional $F_{\wtset}[n]$ depends on the set of weights $\wt_m$. Ref. \onlinecite{GOK88} introduced the following set of weights, so that only one parameter $\wt$ is needed:
\ben
\wt_m=\left\{\begin{array}{ll}
\frac{1-\wt g_I}{M_I-g_I} & m \le M_I-g_I,\\
\wt & m>M_I-g_I,
\end{array}\right.
\label{eqn:theory:OGKens}
\een
where $\wt\in[0,1/M_I]$. In this ensemble, here called GOK, $I$ denotes the set of degenerate states (or `multiplet') with the highest energy in the ensemble, $g_I$ is the multiplicity of the $I$-th multiplet, and $M_I$ is the total number of states up to the $I$-th multiplet. GOK ensembles must contain full sets of degenerate states to be well-defined. The weight parameter $\wt$ interpolates between two ensembles: the equiensemble up to the $I$-th multiplet ($\wt=1/M_I$) and the equiensemble up to the $(I-1)$-th multiplet ($\wt=0$). All previous studies of EDFT have been based on this type of ensemble.

The purpose of EDFT is to calculate excited-state properties, not ensemble properties. With the GOK ensemble, the excitation energy of multiplet $I$ from the ground state, $\omega_I$, is obtained using ensembles up to the $I$-th multiplet as
\ben
\omega_I=\frac{1}{g_I}\left.\frac{\partial E_{I,\wt}}{\partial \wt}\right|_{\wt=\wt_I}+\sum_{i=0}^{I-1}\frac{1}{M_i}\left.\frac{\partial E_{i,\wt}}{\partial \wt}\right|_{\wt=\wt_i},
\label{eqn:theory:exciteng}
\een
which simplifies to
\ben
\omega_1=\omega_{\text{s},1,\wt}+\left.\frac{\partial E\xcw[n]}{\partial \wt}\right|_{n=n_\wt}
\een
for the first excitation energy. Eq. \parref{eqn:theory:exciteng} holds for any valid ${\wt_i}$'s if the ensemble KS systems are exact, despite every term in Eq. \parref{eqn:theory:exciteng} being $\wt$-dependent. No existing $E\xcw$ approximations satisfy this condition.\cite{OGKb88,N98}

Levy\cite{L95} pointed out that there is a special case for $\wt\to0$ of bi-ensembles ($I=2$, with all degenerate states within a multiplet having the same density),
\ben
\begin{split}
\Delta v\xc&=\lim_{\wt\to0}\left.\frac{\partial E\xcw[n]}{\partial \wt}\right|_{n=n_\wt}\\
&=\left[\lim_{\wt\to0}v\xcw[n_\wt](\br)\right]-v_{\sss{xc},\wt=0}[n_{\wt=0}](\br)
\end{split}
\label{eqn:theory:dervdisc}
\een
for finite $r$, where $\Delta v\xc$ is the change in the KS highest-occupied-molecular-orbital (HOMO) energy between $\wt=0$ (ground state) and $\wt\to0_+$.\cite{note1} $\Delta v\xc$ is a property of electron-number-neutral excitations, and should not be confused with the ground-state derivative discontinuity $\Delta\xc$, which is related to ionization energies and electron affinities.\cite{DG90}

\subsection{Degeneracies in the Kohn-Sham system}
\label{sect:theory:connection}
\begin{figure}[htbp]
\includegraphics[width=\columnwidth]{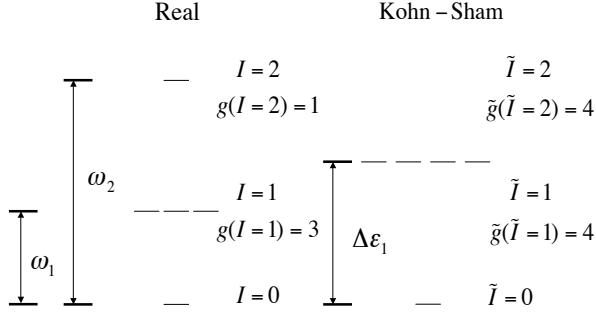}
\caption{Diagram of the interacting and KS multiplicity structure for He. Degeneracy of the $I$-th multiplet is $g(I)$; tildes denote KS values. For instance, $\tilde{I}=2$ refers to the KS multiplet used to construct the second (singlet) multiplet of the real system ($I=2$), as is described in Sec. \ref{sect:theory:correctvHx}.}
\label{fig:cartoon}
\end{figure}

Taking the He atom as our example, the interacting system has a non-degenerate ground state, triply degenerate first excited state, and a non-degenerate second excited state. However, the KS system has a four-fold degenerate first excited state (corresponding to four Slater determinants), due to the KS singlet and triplet being degenerate (Fig. \ref{fig:cartoon}). Consider an ensemble of these states with arbitrary weights. Represent the ensemble energy functional Eq. \parref{eqn:theory:EensDef} as the KS ensemble energy, $E\swset$, plus a correction, $G_\wtset$. The correction then must encode the switch from depending only on the sum of the weights of the excited states as a whole in the KS case to depending on the sum of triplet weights and the singlet weight separately.

For the interacting system, the ensemble energy and density take the forms
\ben
\begin{split}
E_\wtset&=\left(1-\wt_{\sss T}-\wt_{\sss S}\right)E_0+\wt_{\sss T}\omega_1+\wt_{\sss S} \omega_2,\\
n_\wtset(\br)&=\left(1-\wt_{\sss T}-\wt_{\sss S}\right)n_0(\br)+\wt_{\sss T}\Delta n_1(\br)+\wt_{\sss S} \Delta n_2(\br),
\end{split}
\een
where $\omega_i=E_i-E_0$, and so on, $\wt_{\sss T}$ is the sum of the triplet weights, and $\wt_{\sss S}$ is the singlet weight. On the other hand, for the KS system we have
\ben
\begin{split}
E\swset&=\left(1-\wt_{\sss T}-\wt_{\sss S}\right)E_{\sss{S},0}+\left(\wt_{\sss T}+\wt_{\sss S}\right)\Delta \epsilon_{1,\wt},\\
n_\wtset(\br)&=2\left(1-\wt_{\sss T}-\wt_{\sss S}\right)\abs{\phi_{1s}}^2+\left(\wt_{\sss T}+\wt_{\sss S}\right)\left(\abs{\phi_{2s}}^2-\abs{\phi_{1s}}^2\right).
\end{split}
\een
The functional $G_\wtset=E_\wtset-E\swset$ in this case is
\begin{multline}
G_\wtset[n_{\wtset}]=\left(1-\wt_{\sss T}-\wt_{\sss S}\right)\left(E_0-E_{s,0}\right)\\
+\wt_{\sss T}\left(\omega_1-\Delta \epsilon_1\right)+\wt_{\sss S} \left(\omega_2-\Delta \epsilon_1\right),
\end{multline}
showing that the exact ensemble energy functional (which can also be decomposed as in Eq. \parref{eqn:theory:Edecomp}) has to encode the change in the multiplet structure between non-interacting and interacting systems, even for a simple system like the He atom. Such information is unknown \textit{a priori} for general systems, and can be very difficult to incorporate into approximations. This problem can be alleviated if the degeneracies are the result of symmetry. This will be discussed in Sec. \ref{sect:theory:ext:sym}.

\subsection{Approximations}
Available approximations to the ensemble $E\xc$
include the quasi-local-density approximation (qLDA)
functional\cite{K86,OGKb88} and the `ghost'-corrected exact exchange (EXX)
functional.\cite{N98,GPG02} The qLDA functional is based on the
equiensemble qLDA,\cite{K86} and it interpolates between two
consecutive equiensembles:\cite{OGKb88}
\begin{multline}
E_{\sss{XC},I,\wt}^\text{qLDA}[n](\br)=(1-M_I \wt)E_{\sss{XC},I-1}^\text{eqLDA}[n](\br)\\
+M_I \wt E_{\sss{XC},I}^\text{eqLDA}[n],
\end{multline}
where $v\xc^\text{eqLDA}(\br)$ is the
equiensemble qLDA functional defined in terms of
finite-temperature LDA in Ref. \onlinecite{K86}.

The ensemble Hartree energy is defined analogously
to the ground-state Hartree energy as shown in Eq. \parref{eqn:theory:Edecomp}.
Similarly, Nagy provides a
definition of the exchange energy for bi-ensembles:\cite{N98}
\ben
E\xw^\text{Nagy}[n_\uparrow,n_\downarrow]=-\frac{1}{2}\sum_\sigma\int\intd^3r\intd^3r'\;\frac{\abs{n_\sigma(\br,\br')}^2}{\abs{\br-\br'}},
\label{eqn:theory:ExNagy}
\een
where $n_\sigma(\br,\br')$ is the reduced density matrix defined analogously to its
ground-state counterpart, assuming a spin-up electron is
excited in the first excited state:
\ben
n_{\sigma,\wt}(\br,\br')=\sum_{j=1}^{N_\sigma}n_{j,\sigma}(\br,\br')+\delta_{\sigma,\uparrow}\wt\left(n_{\text{L}\uparrow}(\br,\br')-n_{\text{H}\uparrow}(\br,\br')\right),
\label{eqn:theory:1dmNagy}
\een
with $n_{j,\sigma}(\br,\br')=\phi_{j,\sigma}(\br)\phi^{*}_{j,\sigma}(\br')$, \mbox{$L$$\uparrow$}$=N_\uparrow +1$ and \mbox{$H$$\uparrow$}$=N_\uparrow$, the spin-up lowest-unoccupied-molecular-orbital (LUMO) and HOMO, respectively.  Both $E\H$ in Eq. \parref{eqn:theory:Edecomp} and \parref{eqn:theory:ExNagy} contain `ghost' terms,\cite{GPG02} which are cross-terms between different states in the ensemble due to the summation form of $n_\wt(\br)$ in Eq. \parref{eqn:theory:den} and $n_\wt(\br,\br')$ in Eq. \parref{eqn:theory:1dmNagy}. An EXX functional is obtained after such spurious terms are corrected. For two-state ensembles, the GPG XC energy functional is then
\begin{align}
E\xw^{\text{GPG}}&[n_\uparrow,n_\downarrow]=\int\frac{\intd^3r\intd^3r'}{\abs{\br-\br'}}\left\{-\frac{1}{2}\left(n_\sigma(\br,\br')\right)^2\right.\notag\\
&\left.+\wt\wtb\left[n_{\text{H}\uparrow}(\br,\br')n_{\text{L}\uparrow}(\br,\br')-n_{\text{H}\uparrow}(\br')n_{\text{H}\uparrow}(\br')\right]\right\},
\end{align}
where $\wtb=1-\wt$. These `ghost' corrections are small compared to the Hartree and exchange energies. However, they are large corrections to the excitation energies, as Eq. \parref{eqn:theory:exciteng} contains energy derivatives instead of energies. Table \ref{table:atomexx} shows a few examples.

With the help of the exact ensemble KS systems to be presented in this paper, we notice inconsistencies with the GPG functional. These problems and our proposed solutions will be explained in Secs. \ref{sect:theory:Hartree} and \ref{sect:theory:correctvHx}.

\section{Theoretical considerations}
\label{sect:theory:mod}
In this section, we extend EDFT to improve the consistency and generality of the theory.

\subsection{Choice of Hartree energy}
\label{sect:theory:Hartree}
The energy decomposition in Eq. \parref{eqn:theory:Edecomp} is analogous
to its ground-state counterpart. 
However, unlike $T\s$ and $V$, the choices for $E\H$ and $E\x$ and $E\c$ are ambiguous;
only their sum is uniquely determined. As shown
in Eq. \parref{eqn:theory:Edecomp} and \parref{eqn:theory:ExNagy},
definitions for $E\H$ and $E\x$ can introduce `ghost' terms. 
Corrections can be considered either a part of
$E\H$ and $E\x$ or a part of $E\c$. 
Such correction terms also take a complicated form when generalized to multi-state ensembles.

A more natural way of defining $E\H$ and $E\xc$ for ensembles can
be achieved by considering the purpose of this otherwise arbitrary energy decomposition. 
In the ground-state case, the electron-electron repulsion reduces\cite{LS73} to the Hartree
energy for large $Z$, which is a simple functional of the density.
The remaining unknown, $E\xc$ (and its components $E\x$ and $E\c$), is a small
portion of the total energy, so errors introduced by
approximations to it are small. 

For ensembles, 
we propose a slightly different energy decomposition. 
Instead of defining $E\H$ and $E\x$ in analogy to their ground-state counterparts, 
we first define the combined Hartree-exchange energy $E\Hx$, which is
the more fundamental object in EDFT. $E\Hx$ can be explicitly represented as the trace of the KS density matrix:
\ben
E\Hxwset=\tr\{\hat{D}\swset\hat{V}\ee\}=\sum_{m=0}^M \wt_m\matelem{\Phi_m}{\hat{V}\ee}{\Phi_m}.
\label{eqn:EHxdef}
\een
For the ground state, both Hartree and exchange contributions are first-order in the adiabatic coupling constant,
while correlation consists of all higher-order terms.  According to the definition above, we retain this
property in the ensemble.
Eq. \parref{eqn:EHxdef} contains no `ghost' terms by definition, eliminating the need to correct them.
As a consequence, the correlation energy, $E\c$, is defined and decomposed as
\ben
E\cwset=E\Hxcwset-E\Hxwset=T\cwset+U\cwset,
\een
where $E\Hxcwset=E_{\wtset}-T\swset-V$, $T\cwset=T_{\wtset}-T\swset$ and $U\cwset=E\cwset-T\cwset$.

This form of $E\Hx$ reveals a deeper problem in EDFT. As demonstrated in
Sec. \ref{sect:theory:connection}, the multiplet structure of real and KS
He atoms is different. Real He has a triplet state and a singlet state
as the first and second excited states, but KS He has four degenerate single Slater determinants as
the first excited states. Worse, the KS single Slater determinants are not 
eigenstates of the total spin operator $\hat{S}^2$, so their ordering is completely arbitrary. The KS system is constructed to yield only the real spin densities, not other quantities. KS wavefunctions that are not eigenstates of $\hat{S}^2$ do not generally affect commonly calculated ground-state DFT properties,\cite{PRCS09} but things are clearly different in EDFT. Consider the bi-ensemble of the ground state and the triplet excited state of He. Then 
$E\Hxw[\n_\wt]$ depends on which three of the four KS excited-state Slater determinants are chosen, 
though it must be uniquely defined. Therefore, we choose the KS wavefunctions in EDFT to
be linear combinations of the degenerate KS Slater determinants,
preserving spatial and spin symmetries and eliminating ambiguity in $E\Hx$. 
Such multi-determinant, spin eigenstates are also required for construction
of symmetry-projected ensembles, as described in Sec. \ref{sect:theory:ext:sym}.

The multi-determinant KS eigenstates and ensemble $E\Hx$ proposed here
avoid the errors in `ghost'-corrected EXX,\cite{GPG02} which introduces
spurious spin-polarization in closed-shell systems and inherent ambiguity in the
treatment of triplet states. 
We observe considerable improvement in the first excitation energies
of some atoms, as reported in the fourth line of Table \ref{table:atomexx}.

With $E\Hx$ fixed, the definitions of $E\H$ and $E\x$ depend on one another,
but $E\c$ does not. Defining a Hartree functional
in the same form as the ground-state
\ben
U[n]=\frac{1}{2}\int\intd^3r\;\int\intd^3r'\;\frac{n(\br)n(\br)}{\abs{\br-\br'}},
\een
we can examine different definitions for the GOK ensemble. We can define a `ghost'-free ensemble
Hartree, $E\H^\text{ens}$, as
\ben
E\Hw^\text{ens}=\sum_{m=0}^M \wt_m U[n_m],
\label{eqn:EHnew}
\een
i.e., the ensemble sum of the Hartree energies of the interacting densities,
or the slightly different
\ben
E\Hw^\text{KS ens}=\sum_{m=0}^M \wt_m U[n_{\text{s},m}],
\label{eqn:EHnew2}
\een
i.e., the ensemble sum of the Hartree energies of the KS densities.
The traditional Hartree definition, 
\ben
E^{\rm trad}\Hw=U[n_{\wt}],
\label{eqn:EHtrad}
\een
introduces `ghost' terms through the fictitious interaction of ground- and
excited-state densities. Traditional and ensemble definitions differ in their
production of `ghosts,' as well as in their $\wt$-dependence. 
The `ghost'-corrected $E\H$ in Ref. \onlinecite{GPG02}
\ben
E\Hw^\text{GPG}=\sum_{m=0}^M \wt_m^2 U[n_{\text{s},m}]
\een
has a different form from Eq. \parref{eqn:EHnew2}, which is
also `ghost'-free. Each of these definitions of $E\H$ reduces to the ground-state
$E\H$ when $\wt_0=1$ and satisfies simple inequalities such as $E\H>0$ and $E\x<0$. 
However, this ambiguity in the definition of $E\H$ requires that
an approximated ensemble $E\xc$ be explicit about its compatible $E\H$ definition.

\begin{figure}[htbp]
\includegraphics[height=\columnwidth,angle=-90]{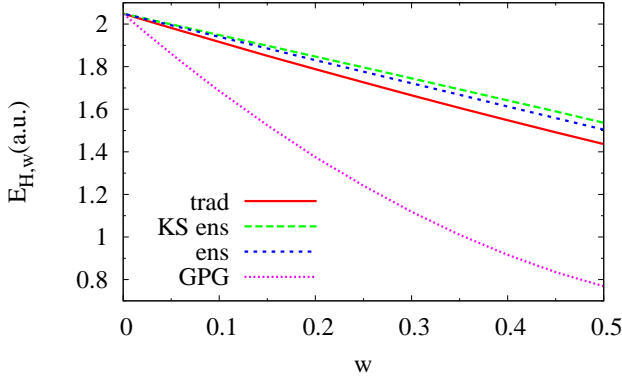}
\caption{Behaviors of the different ensemble Hartree energy definitions for the singlet ensemble of He.}
\label{fig:He:singlet:EH}
\end{figure}

The different flavors of $E\Hw$ are compared for the He singlet ensemble\cite{YTPB14} in
Fig. \ref{fig:He:singlet:EH}. Even though $E\Hw^\text{ens}$ and $E\Hw^\text{KS ens}$ do
not contain `ghost' terms by definition, their magnitude is slightly bigger than
that of $E_{\text{H}}^\text{trad}$, which is not `ghost'-free. This apparent contradiction stems from 
$E\Hw^\text{ens}$ and $E\Hw^\text{KS ens}$ depending linearly on $\wt$, while $E\Hw^\text{trad}$
depends on $\wt$ quadratically. The quadratic dependence on $\wt$ is made explicit with 
the `ghost'-corrected $E\Hw^{\text{GPG}}$ of Ref. \onlinecite{GPG02}. 
Comparing with the `ghost'-free $E\Hw^\text{ens}$ and $E\Hw^\text{KS ens}$, 
it is clear that $E\H^\text{GPG}$ overcorrects in a sense, and 
is compensated by an over-correction of the opposite direction in $E\x^\text{GPG}$ (see Supplemental Material).

The traditional definition of Eq. \parref{eqn:EHtrad} has the advantage that $v\H(\br)$
is a simple functional derivative with respect to the ensemble density. 
Any other definition requires solving an optimized effective potential (OEP)\cite{SH53,TS76}-type
equation to obtain $v\H$. On the other hand, an approximated $E\xc$ 
compatible with $E\H^\text{trad}$ requires users to approximate the
corresponding `ghost' correction as part of $E\xc$. 
Since the ghost correction is usually non-negligible, this is a major source of error for the qLDA functional.

\subsection{Symmetry-eigenstate Hartree-exchange (SEHX)}
\label{sect:theory:correctvHx}
As mentioned previously, the `ghost'-corrected EXX of Ref. \onlinecite{GPG02} introduces
spurious spin-polarization even for closed-shell systems. The root of this problem is the
use of single-Slater-determinant wavefunctions, which are not symmetry eigenstates. 
We have now identified $E\Hx$ as being more consistent with the EDFT formalism than $E\H$ and $E\x$. 
Having also justified multi-determinant ensemble KS wavefunctions, we now derive 
a spin-consistent EXX potential, the symmetry-eigenstate Hartree-exchange (SEHX).  
Define the two-electron repulsion integral
\ben
\left(\mu\nu\mid\kappa\lambda\right)=\int\frac{\intd^3r\intd^3r'}{\abs{\br-\br'}}\phi^*_\mu(\br)\phi^*_\nu(\br')\phi_\kappa(\br)\phi_\lambda(\br')
\een
and
\ben
L_{\mu\nu\kappa\lambda}=\left(\mu\nu\mid\kappa\lambda\right)\delta_{\sigma_\mu,\sigma_\kappa}\delta_{\sigma_\nu,\sigma_\lambda}.
\een
$\phi_\mu(\br)$ denotes the $\mu$-th KS orbital and $\sigma_\mu$ its spin state.  If the occupation of the $p$-th Slater determinant of the $\mu$-th KS orbital of the $\tilde{i}$-th multiplet of the exact system is $f_{p\mu}^{(\tilde{i})}$, define
\ben
\alpha_{\mu,\nu}^{(i,k)}=\sum_{p=1}^{\tilde{g}(\tilde{i})}C_p^{(i,k)}f_{p,\mu}^{(\tilde{i})}f_{p,\nu}^{(\tilde{i})},
\een
where $\tilde{g}(\tilde{i})$ is the KS multiplicity of the $i$-th multiplet, and $C$'s are the coefficients of the multi-determinant wavefunctions defined by
\ben
\Psi_{\text{s}}^{(i,k)}(\br_1,\ldots,\br_N)=\sum_{p=1}^{\tilde{g}(\tilde{i})}C^{(i,k)}_{p}\tilde{\Psi}_{\text{s},p}^{\tilde{i}}(\br_1,\ldots,\br_N).
\een
$\tilde{\Psi}_\text{s}$ is a KS single Slater determinant. Note the numbering of the KS multiplets, $\tilde{i}$, depends on $i$, the numbering of the exact multiplet structure. The $C$ coefficients are chosen according to the spatial and spin symmetries of the exact state.  Now, with $p$ and $q$ KS single Slater determinants of the KS multiplet, define
\begin{align}
h^{(i,k)}_{\mu\nu\kappa\lambda}&=\left(\alpha^{(i,k)}_{\mu,\nu}\alpha^{(i,k)}_{\kappa,\lambda}-\sum_{q=1}^{\tilde{g}(\tilde{i})}\left(C_q^{(i,k)}\right)^2f_{q,\mu}^{(\tilde{i})}f_{q,\nu}^{(\tilde{i})}f_{q,\kappa}^{(\tilde{i})}f_{q,\lambda}^{(\tilde{i})}\right)\notag\\
&\times\prod_{\eta\neq\mu,\nu,\kappa,\lambda}^{\tilde{g}(\tilde{i})}\delta_{f_{p,\eta}^{\tilde{i}},f_{q,\eta}^{\tilde{i}}}, 
\end{align}
in order to write
\ben
H^{(i,k)}=\sum_{\substack{\mu,\nu>\mu \\ \kappa,\lambda>\kappa}}\left(L_{\mu\nu\kappa\lambda}-L_{\mu\nu\lambda\kappa}\right)h^{(i,k)}_{\mu\nu\kappa\lambda}.
\een
Then, if 
\ben
G^{(i,k)}=\sum_{\mu,\nu>\mu}\left(L_{\mu\mu\nu\nu}-\Re{L_{\mu\nu\mu\nu}}\right)\sum_{p=1}^{\tilde{g}(\tilde{i})}\abs{C_p^{(i,k)}}^2f^{\tilde{i}}_{p,\mu}f^{\tilde{i}}_{p,\nu},
\een
the Hartree-exchange energy for up to the $I$-th multiplet is
\ben
E\Hxwset^\text{SEHX}=\sum_{i=1}^I\sum_{k=1}^{g(i)}\wt^{(i,k)}\left\{G^{(i,k)}+H^{(i,k)}\right\},
\een
where $g(i)$ is the exact multiplicity of the $i$-th multiplet. The $v\Hxwset$ potential is then
\ben
\begin{split}
v_{\sss{HX},\wtset,\sigma}^\text{SEHX}(\br)&=\frac{\delta E\Hxwset}{\delta n_{\wtset,\sigma}(\br)}\\
&=\int\intd^3r'\;\sum_j\frac{\delta E\Hxwset}{\delta \phi_{j,\sigma}(\br')}\frac{\delta \phi_{j,\sigma}(\br')}{\delta n_{\wtset,\sigma}(\br)}+\text{c.c.},
\end{split}
\label{eqn:theory:vHxgeneral}
\een
which yields an OEP-type equation for $v\Hxwset(\br)$.

The $v\Hxwset(\br)$ of Eq. \parref{eqn:theory:vHxgeneral} produces neither `ghost' terms nor spurious spin-polarizations. For closed-shell systems, Eq. \parref{eqn:theory:vHxgeneral} yields $v_{\sss{HX},\wtset,\uparrow}(\br)=v_{\sss{HX},\wtset,\downarrow}(\br)$, unlike Ref. \onlinecite{GPG02}. An explicit $v\Hxwset(\br)$ can be obtained by applying the usual Krieger-Li-Iafrate(KLI)\cite{KLI90} approximation. Here we provide the example of the singlet bi-ensemble studied in our previous paper.\cite{YTPB14} $E\Hx$ for a closed-shell, singlet ensemble is
\begin{multline}
E\Hxw^\text{SEHX}=\int\frac{\intd^3r\intd^3r'}{\abs{\br-\br'}}\left\{n^\text{orb}_1(\br)n^\text{orb}_1(\br')\right.\\
+\wt\left[n^\text{orb}_1(\br)\left(n^\text{orb}_2(\br')-n^\text{orb}_1(\br')\right)\right.\\
\left.\left.+\phi_1^*(\br)\phi_2^*(\br')\phi_1(\br')\phi_2(\br)\right]\right\},
\end{multline}
where $n^\text{orb}_j(\br)=\abs{\phi_j(\br)}^2$ is the KS orbital density. Spin is not explicitly written out because the system is closed-shell. After applying the KLI approximation, we obtain
\begin{multline}
v\Hxw(\br)=\frac{1}{2n_\wt(\br)}\left\{(2-\wt)n^\text{orb}_1(\br)\left[v_1(\br)+\bar{v}_{\sss{HX}1}-\bar{v}_1\right]\right.\\
\left.+\wt~n^\text{orb}_2(\br)\left[v_2(\br)+\bar{v}_{\sss{HX}2}-\bar{v}_2\right]\right\},
\label{eqn:theory:correctvHx}
\end{multline}
with
\begin{multline}
v_1(\br)=\frac{1}{(2-\wt)}\int\frac{\intd^3r'}{\abs{\br-\br'}}\left[2(1-\wt)n^\text{orb}_1(\br')\right.\\
\left.+\wt\left(n^\text{orb}_2(\br')+\phi_1^*(\br')\phi_2^*(\br)\phi_2(\br')/\phi_1^*(\br)\right)\right],
\end{multline}
\ben
v_2(\br)=\int\frac{\intd^3r'}{\abs{\br-\br'}}\left[n^\text{orb}_1(\br')+\frac{\phi_1^*(\br)\phi_2^*(\br')\phi_1(\br')}{\phi_2^*(\br)}\right],
\een
and
\ben
\bar{v}_j=\int\intd^3r\;v_j(\br)n^\text{orb}_j(\br).
\een
Eq. \parref{eqn:theory:correctvHx} is an integral equation for $v\Hx(\br)$ that can be easily solved.

To fully understand the performance of $v\Hx(\br)$, self-consistent EDFT calculations would be needed at different values of $\wt$, which is beyond the scope of this paper. We demonstrate the performance at $\wt=0$ later in Sec. \ref{sect:gsatoms}.

\subsection{Symmetry-projected Hamiltonian}
\label{sect:theory:ext:sym}
The ensemble variational principle holds for any Hamiltonian. If the Hamiltonian $\hat{H}$ commutes with another operator $\hat{O}$, one can apply to $\hat{H}$ a projection operator formed by the eigenvectors of $\hat{O}$. One obtains a new Hamiltonian, and the ensemble variational principle holds for this subspace of $\hat{H}$, allowing an EDFT to be formulated.

An example would be the total spin operator $\hat{S}^2$, where
\ben
S^2=\sum_{S=0}^{\infty}(2S+1)\ket{S}\bra{S}
\een
and $\ket{S}$ are its eigenvectors. Define a new Hamiltonian $\hat{H}_1$ as
\ben
\hat{H}_1=\ket{S}\bra{S}\hat{H}.
\een
$\hat{H}_1$ has the same set of eigenvectors as $\hat{H}$, but the eigenvalues are 0 for the eigenvectors not having spin $S$. Since one can change the additive constant in $\hat{H}$ arbitrarily, it is always possible to make the eigenvalues of any set of spin-$S$ eigenvectors negative and thus ensure that they are the lowest energy states of $\hat{H}_1$. The ensemble variational principle holds for ensembles of spin-$S$ states. We have employed this symmetry argument in our previous paper\cite{YTPB14} for a purely singlet two-state ensemble of the He atom.

A similar statement is available in ground-state DFT, allowing direct calculation of the lowest state of a certain symmetry.\cite{Gunnarsson1976,Goerling1993} The differences between the subspace and full treatments are encoded in the differences in their corresponding $E\xc$. Thus the lowest two states within each spatial and spin symmetry category can be treated in EDFT in a two-state-ensemble fashion, which is vastly simpler than the multi-state formalism.

Since the multiplet structures of the interacting system and the KS system must be compatible, a symmetry-projected ensemble also requires a symmetry-projected KS system, which is impossible if KS wavefunctions are single Slater determinants, as discussed in Sec. \ref{sect:theory:Hartree}.

\section{Exact conditions}
\label{sect:theory:ext}
Here we prove some basic relations for the signs of various components of the KS scheme and a virial for the potentials.  We describe a feature of the ensemble derivative discontinuity and extraction of excited properties from the ground state.

\subsection{Inequalities}
\label{sect:theory:inequ}
Simple exact inequalities of the energy components (such as $E\c<0$) have been proven in ground-state DFT.\cite{DG90} If these are true in EDFT, experiences designing approximated $E\xc$ in ground-state DFT may be transferrable to EDFT. Here we show that inequalities related to the correlation energy are still valid in EDFT.

Due to the variational principle,\cite{GOKb88} the wavefunctions that minimize the ensemble energy Eq. \parref{eqn:theory:EensDef} are the interacting wavefunctions $\Psi_m$. Thus
\ben
E\cwset=\tr\{\hat{D}_{\wtset}\hat{H}\}-\tr\{\hat{D}\swset\hat{H}\}\le 0.
\label{eqn:theory:Ecineq}
\een
The existence of a non-interacting KS system\cite{GOK88} means $T\swset$ is the smallest possible kinetic energy for a given density $n_\wtset(\br)$, resulting in
\ben
T\cwset=T_{\wtset}-T\swset\ge 0.
\label{eqn:theory:Tcineq}
\een
From Eq. \parref{eqn:theory:Ecineq} and \parref{eqn:theory:Tcineq} we immediately obtain
\ben
U\cwset=E\cwset-T\cwset\le 0,
\een
and
\ben
\abs{U\cwset}\ge\abs{T\cwset}.
\een
These inequalities are later verified with exact ensemble KS calculations.

\subsection{Virial Theorem}
Since EDFT is a variational method, one expects that the virial theorem holds. A brief argument was given in Ref. \onlinecite{YTPB14}, but we provide a straightforward proof here. We apply the usual coordinate scaling on the wavefunctions:\cite{PY89}
\ben
\Psi_{m,\gamma}(\br_1,\ldots,\br_N)=\gamma^{3N/2}\Psi_m(\gamma\br_1,\ldots,\gamma\br_N).
\een
According to the variational principle, the exact interacting wavefunctions $\Psi_m$ minimize the ensemble energy $E_{\wtset}$ for a given set $\wtset$. First-order variations of the ensemble energy therefore vanish. Thus
\ben
\left.\frac{d}{d\gamma}\tr\{\hat{D}_{\wtset,\gamma}\hat{H}\}\right|_{\gamma=1}=0,
\een
where $\hat{D}_{\wtset,\gamma}=\sum_{m=0}^M \wt_m\ket{\Psi_{m,\gamma}}\bra{\Psi_{m,\gamma}}$. Since $\hat{H}=\hat{T}+\hat{V}_{\sss tot}=\hat{T}+\hat{V}+\hat{V}\ee$, we have
\ben
\left.\frac{d}{d\gamma}\tr\{\hat{D}_{\wtset}\}(\gamma^2\hat{T}+\gamma\hat{V}_{\sss tot})\}\right|_{\gamma=1}=0,
\een
yielding
\ben
2T_{\wtset}[n_{\wtset}]+V_{\text{ee},\wtset}[n_{\wtset}]=\int\intd^3r\;n_{\wtset}(\br)\br\cdot\nabla v(\br).
\label{eqn:theory:virial:general}
\een

We can write the energy of an ensemble for a given set of weights $\wtset$ and a given coupling constant $\lambda$ as
\begin{equation}
E_{\wtset}^\lambda[n]=F_{\wtset}^\lambda[n]+\int\intd^3r\;n(\br)v_{\wtset}^\lambda(\br),
\label{eqn:theory:virial:E}
\end{equation}
where $v_{\wtset}^\lambda(\br)$ is the $\wtset$-dependent, one-body potential maintaining a constant density for all degrees of interaction.\cite{LP85} $v_{\wtset}^{\lambda=0}(\br)=v\swset(\br)$, and $v_{\wtset}^{\lambda=1}(\br)=v(\br)$. The functional derivative of Eq. \parref{eqn:theory:virial:E} with respect to $n(\br)$ is
\begin{equation}
\frac{\delta E_{\wtset}^\lambda}{\delta n(\br)}=\frac{\delta F_{\wtset}^\lambda}{\delta n(\br)}+v_{\wtset}^\lambda(\br).
\label{eqn:theory:virial:mu}
\end{equation}
Thus
\begin{equation}
-\int\intd^3r\;n(\br) \br\cdot\nabla\left[\frac{\delta F_{\wtset}^\lambda}{\delta n(\br)}\right]=\int\intd^3r\;n(\br) \br\cdot\nabla v_{\wtset}^\lambda(\br).
\label{eqn:theory:virial:int}
\end{equation}
Applying the ensemble virial theorem of Eq. \parref{eqn:theory:virial:general} to Eq. \parref{eqn:theory:virial:int} yields
\begin{equation}
-\int\intd^3r\;n(\br)\br\cdot\nabla\left[\frac{\delta F_\wtset^\lambda}{\delta n(\br)}\right]=F_\wtset^\lambda[n]+T_\wtset^\lambda[n].
\end{equation}

Considering the KS quantities, we can insert the energy in terms of Hartree-exchange-correlation and the $\lambda$-dependent one-body potential into the general virial:
\begin{equation}
2T_{\wtset}^\lambda[n]=-\lambda E\Hxcwset^\lambda[n]+\lambda T\cwset^\lambda[n]+\int d^3r\;n(\br)\br\cdot\nabla v_{\wtset}^\lambda(\br).
\end{equation}
After this, set $\lambda=1$ for the physical system,
\begin{equation}
2T_{\wtset}[n]=-E\Hxcwset[n]+T\cwset[n]+\int d^3r\;n(\br)\br\cdot\nabla v_{\wtset}(\br),
\end{equation}
and $\lambda=0$ for KS,
\begin{equation}
2T\swset[n]=\int d^3r\;n(\br)\br\cdot\nabla v\swset(\br),
\end{equation}
and subtract. This yields
\begin{equation}
T\cwset[n]=-E\Hxcwset[n]-\int d^3r\;n(\br)\br\cdot\nabla v\Hxcwset(\br)
\label{eq:theory:viral:Hxc}
\end{equation}
and
\begin{equation}
E\Hxwset[n]=-\int d^3r\;n(\br)\br\cdot\nabla v\Hxwset(\br).
\end{equation}
Finally, the virial theorem for the correlation energy takes a similar form as in ground-state DFT:
\ben
T\cwset[n]=-E\cwset[n]-\int d^3r\;n(\br)\br\cdot\nabla v\cwset(\br).
\label{eqn:theory:virial:xc}
\een

Energy densities have been important interpretation tools in ground-state DFT, and here we provide similar tools for EDFT. The integrand of Eq. \parref{eq:theory:viral:Hxc} can be interpreted as an energy density, since integrating over all space gives
\ben
\begin{split}
E\Hxcwset+T\cwset&=\int d^3r\left(e\Hxcwset+t\cwset\right)\\
&=-\int d^3r~n(\br)\br\cdot\nabla v\Hxcwset(\br),
\end{split}
\label{eqn:theory:virialengden}
\een
which can easily be converted to an ``unambiguous" energy density.\cite{BCL98}

\subsection{Asymptotic behavior}
Ref. \onlinecite{L95} derived the ensemble derivative discontinuity of Eq. \parref{eqn:theory:dervdisc} for bi-ensembles, in the limit of $\wt\to0$. For finite $\wt$ of an atomic system, as shown in our previous paper,\cite{YTPB14} $\Delta v\xc$ is close to a finite constant for small $r$, and jumps to 0 at some position denoted by $r\c$. We provide the derivation of the location of $r\c$ as a function of $\wt$ here.

For atoms, the HOMO wavefunction and LUMO wavefunctions have the following behavior:
\ben
\begin{split}
\phi_\text{HOMO}(\br)&\sim A r^{\beta}e^{-\alpha r}\\
\phi_\text{LUMO}(\br)&\sim A' r^{\beta'}e^{-\alpha' r},
\end{split}
\een
with $\alpha\geq\alpha'$.  For the bi-ensemble of the ground state and the first excited state, the ensemble density is
\begin{multline}
n_\wt(\br)\sim2\sum_{n=1}^{\text{HOMO}}\abs{\phi_n(\br)}^2\\
+\wt\left(A'^2 r^{2\beta'}e^{-2\alpha' r}-A^2 r^{2\beta}e^{-2\alpha r}\right),\quad r\rightarrow\infty,
\label{eqn:theory:deltavxc:n}
\end{multline}
assuming that the HOMO is doubly-occupied. The behavior of the density at large $r$ is dominated by the density of the doubly-occupied HOMO and the second term. In order to see where the density decay switches from that of the HOMO to the LUMO, we find the $r$-value at which the two differently decaying contributions are equal:
\ben
\left(2-\wt\right)A^2 r^{2\beta}e^{-2\alpha r}=\wt A'^2 r^{2\beta'}e^{-2\alpha r}.
\een
As $\wt\to0$, $r\c$ is then
\ben
r\c\to-\frac{\ln{\wt}}{2\Delta\alpha},
\een
with $\Delta\alpha=\alpha-\alpha'$.

The ionization energies are available for the He ground state and singlet excited state. Since
\ben
n(\br)\sim e^{-2\alpha r}\approx e^{-2\sqrt{2I}r},
\een
we obtain
\ben
r\c\to-0.621 \ln \wt,\quad \wt\to0.
\een
for the He singlet bi-ensemble with $\wt$ close to 0.

\subsection{Connection to ground-state DFT}
\label{sect:gsatoms}
With weights as in Eq. \parref{eqn:theory:OGKens}, calculation of the excitation energies is done
recursively: for the $M$th excited state, one needs to perform an EDFT
calculation with the $M$th state highest in the ensemble, and
another EDFT calculation with the $(M-1)$th as the
highest state, and so on. Thus for the $M$th state, one
needs to perform $M$ separate EDFT calculations for its excitation
energy.

For bi-ensembles, however, the calculation of the excitation energy can be greatly
simplified. Eq. \parref{eqn:theory:exciteng} holds for
$\wt=0$, so one can work with ground-state data only and obtain
the first-excited state energy, without the need for an
explicit EDFT calculation of the two-state ensemble.

We calculate the first excitation energies of various atoms and ions with Eq. \parref{eqn:theory:exciteng} at $\wt=0$ with both qLDA\cite{OGKb88,K86} (based on LDA ground states), EXX,\cite{N98} GPG,\cite{GPG02} and SEHX, with the last three based on OEP-EXX (KLI) ground states.\cite{KLI90} In order to ensure the correct symmetry in the end result, SEHX must be performed on spin-restricted ground states. However, for closed-shell systems, these results coincide with those of spin-unrestricted calculations. We use these readily available results when possible in this paper. The $\wt$-derivatives of the $E\xc$'s for qLDA and EXX required in Eq. \parref{eqn:theory:exciteng} are (considering Eq. \parref{eqn:theory:fine:dExcdw})
\ben
\lim_{\wt\to0}\left.\frac{\partial E\xcw^\text{qLDA}[n]}{\partial \wt}\right|_{n=n_\wt} = M_I \left(E\xc^\text{eqLDA}[I=2,n]-E\xc^\text{LDA}[n]\right),
\een
where $E\xc^\text{LDA}$ is the ground-state LDA functional, and
\begin{eqnarray}
\lefteqn{
\lim_{\wt\to0}\left.\frac{\partial E\xw^{\text{GPG}}[n]}{\partial \wt}\right|_{n=n_\wt} =
\int\!\!\int\frac{\intd^3r\intd^3r'}{\abs{\br-\br'}}}
\nonumber\\
&&
\times\Bigg\{\left[\sum_{j=1}^{N^\uparrow}n_{j}(\br,\br')\right]
\left[n_\text{H}(\br,\br')-n_\text{L}(\br,\br')\right]
\nonumber\\
&&
 -n_\text{H}(\br)n_\text{L}(\br')+n_\text{H}(\br,\br')n_\text{L}(\br,\br')\Bigg\}
\nonumber\\
&&
+\int\intd^3r\;v\xc(\br)[n_\text{H}(\br)-n_\text{L}(\br)],
\label{eqn:atomexx:dExcdw}
\end{eqnarray}
where $j$ sums over the spin-up densities. Only ground state properties are needed to evaluate Eq. \parref{eqn:atomexx:dExcdw}. The results are listed in Table \ref{table:atomexx}. SEHX improves calculated excitation energies for systems where GPG has large errors, such as Be and Mg atoms. 

\begin{table}[t]
\begin{tabular}{cccccccccc}
\hline\hline
\ & He & Li & Li$^+$ & Be & Be$^+$ & Mg & Ca & Ne & Ar\\
\hline
Exp. & 20.62 & 1.85 & 60.76 & 5.28 & 3.96 & 4.34 & 2.94 & 16.7 & 11.6\\
qLDA & - & 1.93 & 53.85 & 3.71 & 4.30 & 3.58 & 1.79 & 14.2 & 10.7\\
EXX & 27.30 & 6.34 & 72.26 & 10.22 & 12.38 & 8.25 & 9.89 & 26.0 & 18.2\\
GPG & 20.67 & 1.84 & 60.40 & 3.53 & 4.00 & 3.25 & 3.25 & 18.2 & 12.1\\
SEHX & 20.67 & 2.08$^*$ & 60.40 & 5.25 & 4.06$^*$ & 4.39 & 3.55 & 18.4 & 12.2\\
\hline\hline
\end{tabular}
\caption{First non-triplet excitation energies (in eV) of various atoms and ions calculated with qLDA, EXX, GPG, and SEHX functionals. qLDA calculations were performed upon LDA (PW92)\cite{PW92} ground states; EXX\cite{N98} ground states were used for the rest. Asterisks indicate use of spin-restricted ground states. qLDA relies on ground-state LDA orbital energy differences; it cannot be used with the single bound orbital of LDA He.}
\label{table:atomexx}
\end{table}

\section{Numerical procedure}
\label{sect:numerical}
We invert the ensemble KS equation with exact densities to obtain the exact KS potential. We describe the numerical inversion procedure in Ref. \onlinecite{YTPB14}. For ease in obtaining the Hartree potential, $E\H$ is always chosen to be $E\H^\text{trad}$. The resulting KS potential, being exact, does not depend on the choice of $E\H$, but $E\xc$ and $v\xc(\br)$ reported in later sections are those compatible with $E\H^\text{trad}$ and $v\H^\text{trad}(\br)$, respectively. For simplicity, only GOK-type ensembles [Eq. \parref{eqn:theory:OGKens}] are considered, though there is no difficulty adapting the method to other types of ensembles. With this numerical procedure, $v\xcw(\br)$ is determined up to an additive constant. 

We implemented the numerical procedure on a real-space grid. The ensemble KS equation (\ref{eqn:theory:KS}) is solved by direct diagonalization of the discrete Hamiltonian. The grid is in general nonuniform, which complicates the discretization of the KS kinetic energy operator. We tested two discretization schemes, details of which are available in the Supplemental Material.  Based on these tests, all results presented in this paper have been obtained using the finite-difference representation
\begin{multline}
-\frac{1}{2}\frac{d^2\phi(x)}{dx^2}\approx\frac{\phi(x_i)}{(x_i-x_{i-1})(x_{i+1}-x_i)}\\
-\frac{\phi(x_{i-1})}{(x_i-x_{i-1})(x_{i+1}-x_{i-1})}-\frac{\phi(x_{i+1})}{(x_{i+1}-x_i)(x_{i+1}-x_{i-1})}.
\end{multline}

\subsection{Derivative Corrections}
Exactness of the inversion process can be verified by calculating the excitation energies with Eq. \parref{eqn:theory:exciteng} at different $\wt$ values. Eq. \parref{eqn:theory:exciteng} requires calculating $E\xcw$ of the exact ensemble KS system,
\begin{multline}
E\xcw[n_\wt]=E_\wt-E_{\sss{s},\wt}\\
+\int\intd^3r\;n_\wt(\br)\left[\frac{v\H[n_\wt](\br)}{2}+v\xcw[n_\wt](\br)\right].
\label{eqn:theory:Exccalc}
\end{multline}

Since we do not have a closed-form expression for the exact $E\xc$, its derivative can only be calculated numerically. However, the numerical derivative of $E\xc$, $\partial E\xcw[n_\wt]/\partial \wt$, is not the quantity required in Eq. \parref{eqn:theory:exciteng}. It is related to the true derivative through \begin{multline}
\left.\frac{\partial E\xcw[n]}{\partial \wt}\right|_{n=n_\wt}=\frac{\partial E\xcw[n_\wt]}{\partial \wt}\\-\int\intd^3r\;v\xcw[n_\wt](\br)\frac{\partial n_\wt(\br)}{\partial \wt}.
\label{eqn:theory:fine:dExcdw}
\end{multline}
The correction to the numerical derivative of $E\xcw$ adjusts for the $\wt$-dependence of the ensemble density, which is not inherent to $E\xcw$. All our calculations show that the two terms on the right hand side of Eq. \parref{eqn:theory:fine:dExcdw} are of the same order of magnitude. This shows that the exact $E\xcw[n]$ changes more slowly than $n_\wt(\br)$ as $\wt$ changes. Though the calculations of $E\xcw$ and $\partial E\xcw[n]/\partial \wt|_{n=n_\wt}$ both involve integrations containing $v\xcw(\br)$, they are independent of the additive constant.

\section{Results}
\label{sect:results}
We apply the numerical procedure described in Sect. \ref{sect:numerical} to both 1D and 3D model systems in order to further demonstrate our method for inverting ensemble densities.

\subsection{1D flat box}
\label{sect:1dbox}
The external potential of the 1D flat box is
\ben
v(x)=\left\{\begin{array}{cl}0, & 0<x<L,\\ \infty, & x\le 0\text{ or }x\ge L. \end{array}\right.
\een
The exact wavefunctions can be solved numerically for two electrons with the following soft-Coulomb interaction:
\ben
v_{\sss{SC}}(x,x')=\frac{1}{\sqrt{(x-x')^2+a^2}},
\label{eqn:1dbox:softCoulomb}
\een
where we choose $a=0.1$.

\begin{table}[htbp]
\begin{tabular}{cccc}
\hline\hline
$I$ & $E$ & $T$\\
\hline
$0$ (singlet) & 15.1226 & 10.0274\\
$1$ (triplet) & 27.5626 & 24.7045\\
$2$ (singlet) & 30.7427 & 24.7696\\
$3$ (singlet) & 43.9787 & 39.6153\\
$4$ (triplet) & 52.8253 & 49.3746\\
\hline\hline
\end{tabular}
\caption{Total and kinetic energies in a.u. for a unit-width box, including a doubly-excited state ($I=3$).}
\label{table:boxET}
\end{table}
Table \ref{table:boxET} shows the total and kinetic energies of the exact ground state and first four excited states for $L=1$ a.u., calculated on a 2D uniform grid with 1000 points for each position variable. The third excited state is a doubly-excited state corresponding to both electrons  occupying the second orbital of the box. Fig. \ref{fig:1dbox:general} shows the exact densities of the ground state and first four excited states, together with the XC potential of equiensembles containing 1 to 5 multiplets. Table \ref{table:1dbox:multistate} lists calculated excitation energies, showing that the excitation energy is independent of $\wt$, no matter how many states are included in the ensemble. This is a non-trivial exact condition for the ensemble $E\xc$.

\begin{figure}[htbp]
\includegraphics[height=\columnwidth,angle=-90]{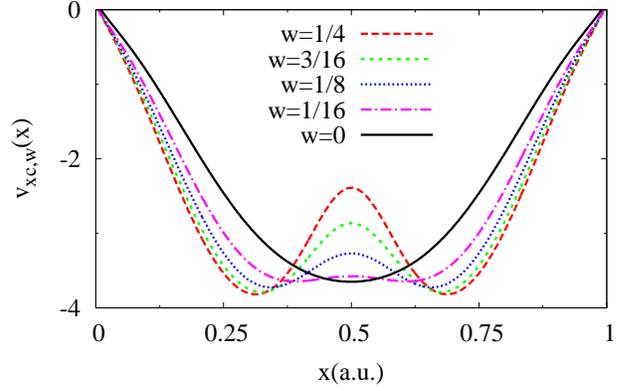}
\caption{Exact ensemble XC potentials of the 1D box with two electrons. The ensemble contains the ground state and the first (triplet) excited state.}
\label{fig:1dbox:vxc}
\end{figure}

Double excitations are generally difficult to calculate. It has been shown that adiabatic TDDFT cannot treat double or multiple excitations.\cite{EGCM11} Table \ref{table:1dbox:multistate} shows that there is no fundamental difficulty in treating double excitations with EDFT. Fig. \ref{fig:1dbox:general} shows that $v\xcw(\br)$ for the 4-multiplet equiensemble resembles the potentials of other ensembles. The exact two-multiplet ensemble XC potentials at different $\wt$ are plotted in Fig. \ref{fig:1dbox:vxc}.  The bump up near the center of the box in these potentials ensures that the ensemble KS density matches that of the real ensemble density. Increasing the proportion of the excited state density (see Fig. \ref{fig:1dbox:general}) included in the ensemble density requires a corresponding increase in the height of this bump (see Supplemental Material).  With no asymptotic region, there is no derivative discontinuity for the box, and $v_{{\sss XC},\wt\to0}(\br)$ is equal to the ground-state $v\xc(\br)$. Energy components for the bi-ensemble of the 1D box satisfy the inequalities shown in Sec. \ref{sect:theory:inequ} and are reported in the Supplemental Material.

\begin{table}[htbp]
\begin{tabular}{cccc}
\hline\hline
\multicolumn{4}{l}{2-multiplet: $\omega_1=12.4399$ hartree}\\
\hline
$\wt_2$ & 0.25 & 0.125 & 0.03125\\
$E_{1,\wt_2}^\text{KS}-E_{0,\wt_2}^\text{KS}$ & 13.9402 & 13.9201 & 13.8932\\
$\partial E{\sss{xc},\wt_2}[I=2,n]/\partial \wt_2|_{n=n_{\wt_2}}$ & -4.5010 & -4.4407 & -4.3598\\
$(E_1-E_0)_{\wt_2}$ & 12.4399 & 12.4399 & 12.4399\\
\hline
\multicolumn{4}{l}{3-multiplet: $\omega_2=15.6202$ hartree}\\
\hline
$\wt_3$ & 0.2 & 0.1 & 0.025\\
$E_{2,\wt_3}^\text{KS}-E_{0,\wt_3}^\text{KS}$ & 14.2179 & 14.0757 & 13.9735\\
$\partial E{\sss{xc},\wt_3}[I=3,n]/\partial \wt_3|_{n=n_{\wt_3}}$ & 2.7358 & 2.7713 & 2.7969\\
$(E_2-E_0)_{\wt_2,\wt_3}$ & 15.6202 & 15.6201 & 15.6202\\
\hline
\multicolumn{4}{l}{4-multiplet: $\omega_3=28.8561$ hartree (double)}\\
\hline
$\wt_4$ & 0.166666 & 0.083333 & 0.020833\\
$E_{3,\wt_4}^\text{KS}-E_{0,\wt_4}^\text{KS}$ & 28.7534 & 28.5826 & 28.4706\\
$\partial E{\sss{xc},\wt_4}[I=4,n]/\partial \wt_4|_{n=n_{\wt_4}}$ & 1.1061 & 1.1186 & 1.1858\\
$(E_3-E_0)_{\wt_2,\wt_3,\wt_4}$ & 28.8561 & 28.8561 & 28.8561\\
\hline
\multicolumn{4}{l}{5-multiplet: $\omega_4=37.7028$ hartree}\\
\hline
$\wt_5$ & 0.111111 & 0.055555 & 0.013888\\
$E_{4,\wt_5}^\text{KS}-E_{0,\wt_5}^\text{KS}$ & 38.8375 & 38.8602 & 38.8746\\
$\partial E{\sss{xc},\wt_5}[I=5,n]/\partial \wt_5|_{n=n_{\wt_5}}$ & -1.1279 & -1.2205 & -1.2787\\
$(E_4-E_0)_{\wt_2,\wt_3,\wt_4,\wt_5}$ & 37.7028 & 37.7027 & 37.7028\\
\hline\hline
\end{tabular}
\caption{Excitation energies of the 1D box calculated at different $\wt$ values using the exact ensemble KS systems and Eq. \parref{eqn:theory:exciteng}. The double excitation (4-multiplet) shows accuracy comparable to that of the single excitation (2-multiplet). All energies are in Hartree. See Supplemental Material for the full table.}
\label{table:1dbox:multistate}
\end{table}

\subsection{Charge-transfer excitation with 1D box}
\label{sect:1dbox:ct}
Charge-transfer (CT) excitations are difficult to treat with approximate TDDFT, due to the lack of overlap between orbitals.\cite{FRM11} With common approximations, the excitation energy calculated by TDDFT is much smaller than experimental values.\cite{U12} Here we provide a 1D example of an excited state with CT character, showing that there is no fundamental difficulty in treating CT excitations with EDFT. Since EDFT calculations do not involve transition densities, they do not suffer from the lack-of-orbital-overlap problem in TDDFT.

\begin{figure}[htbp]
\includegraphics[height=\columnwidth,angle=-90]{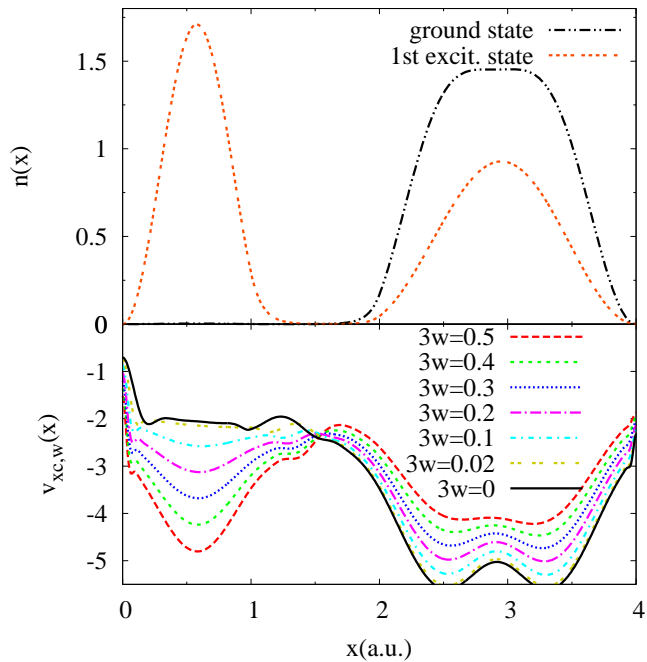}
\caption{Exact densities and ensemble xc potentials of the 1D charge-transfer box.}
\label{fig:1dbox:ct:general}
\end{figure}

The external potential for the CT box is
\ben
v(x)=\left\{
\begin{array}{ll}
0 & x\in[0,1]\cup[2,4]\\
20 & x\in(1,2)\\
\infty & x<0\text{ or }x>4,
\end{array}
\right.
\label{eqn:1dbox:ct:vext}
\een
with the barrier dimensions chosen for numerical stability of the inversion process. The lowest two eigenstate densities are given in the top of Fig. \ref{fig:1dbox:ct:general}. The ground-state and first-excited-state total and kinetic energies of the CT system described are
\begin{equation}
\begin{split}
E_0&=138.254\,\text{eV},\quad T_0=63.4617\,\text{eV (singlet)},\\
E_1&=140.652\,\text{eV},\quad T_1=112.141\,\text{eV (triplet)}.
\end{split}
\end{equation}
This significant increase in kinetic energy together with a small total energy change designate the CT character of the first excited state. The electrons become distributed between the two wells of the potential, instead of being confined in one well.

The ground- and first-excited-state densities and ensemble XC potentials are plotted in Fig. \ref{fig:1dbox:ct:general}. The  potentials show the characteristic step-like structures of charge-transfer excitations, which align the chemical potentials of the two wells.\cite{HG12,HG13} Table \ref{table:1dbox:CT:E} lists the ensemble energies of the CT box. Excitation energies have larger errors than those for the 1D flat box due to greater numerical instability, but they are still accurate to within $0.01$ eV.

\begin{table}[htbp]
\begin{tabular}{cccc}
\hline\hline
$3\wt$ & 0.5 & 0.1 & 0.02\\
\hline
$E_{1,\wt}^\text{KS}-E_{0,\wt}^\text{KS}$ & 2.2048 & 2.4092 & 2.4317\\
$\partial E\xcw[n]/\partial n|_{n=n_\wt}/3$ & 0.1993 & -0.0108 & -0.0334\\
$\omega_{1,\wt}$ & 2.4042 & 2.3983 & 2.3983\\
\hline\hline
\end{tabular}
\caption{First excitation energy and energy decomposition of the two-multiplet ensemble of the 1D charge-transfer box at different $\wt$ values, calculated using Eq. \parref{eqn:theory:exciteng}. All energies are in eV. The exact first excitation energy is $E_1-E_0=2.3983$ eV. See Supplemental Material for additional data.}
\label{table:1dbox:CT:E}
\end{table}

\subsection{Hooke's atom}
Hooke's atom is a popular model system\cite{Laufer1986,Filippi1994} with the following external potential:
\ben
v(\br)=\frac{k}{2}\abs{\br}^2.
\een
For our calculation, $k=1/4$. Though the first excited state has cylindrical symmetry, we use a spherical grid, as it has been shown that the error due to spherical averaging is small.\cite{KP87} As a closed-shell system, the spatial parts, and therefore the densities, of the spin-up and spin-down ensemble KS orbitals have to be the same, so we treat this system as a bi-ensemble.

The magnitude of the external potential of the Hooke's atom is smallest at $r=0$, and becomes larger as $r$ increases. This is completely different from the Coulomb potential of real atoms. Since the electron-electron interaction is still coulombic, $v\xc(\br)$ can be expected to have a $-1/r$ behavior as $r\to\infty$, which is negligibly small compared to $v(\br)$. Combined with a density that decays faster than real atomic densities, $n(\br)\sim\exp(-ar^2)$ versus $n(\br)\sim\exp(-br)$, convergence of the Hooke's atom $v\xc(\br)$ is difficult in the asymptotic region. Additionally, $v\xc(\br)\gg v(\br)$ for small $r$, so larger discretization errors in this region also contribute to poorer inversion performance. Despite these challenges, we still obtain highly accurate excitation energies.

A logarithmic grid with 550 points ranging from $r=10^{-5}\,\text{a.u.}$ to $10\,\text{a.u.}$ is used for all the Hooke's atom calculations. On this grid, the exact ground- and first excited-state energies are
\ben
E_1=54.42\,\text{eV},\quad E_2=64.19\,\text{eV}.
\een
Calculated $\omega_2$ was 9.786 eV for all values of $\wt$ tested (see Supplemental Material). Unlike the He atom and the 1D flat box, the $n_\wt(\br)$ and $v\xcw(\br)$ show little variation with $\wt$ (see Supplemental Material). The second KS orbital of the Hooke's atom is a $p$-type orbital, which has no radial node and a radial shape similar to that of the first KS orbital. Consequently, the changes in the KS and xc potentials are also smaller.

\subsection{He}
\label{sect:results:He}
Using the methods in Ref. \onlinecite{YTPB14}, we employ a Hylleraas expansion of the many-body wavefunction\cite{drake_94} to calculate highly accurate densities of the first few states of the He atom. We report the exact ensemble XC potentials for He singlet ensemble in that paper. Table \ref{table:He:3state} shows accurate excitation energies calculated from mixed symmetry, three-multiplet, and strictly triplet ensembles, demonstrating the versatility of EDFT.
\begin{figure}[htbp]
\includegraphics[height=\columnwidth,angle=-90]{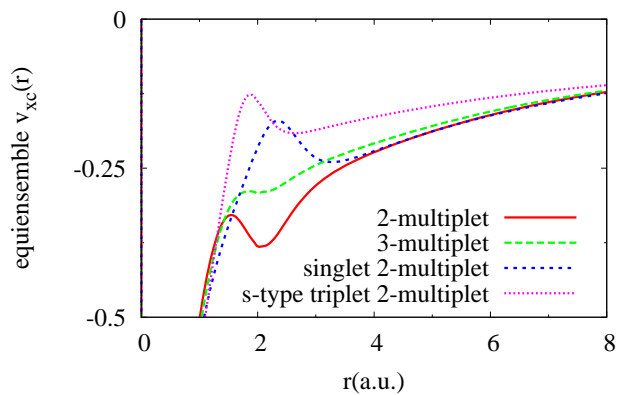}
\caption{Ensemble XC potentials of the singlet-triplet, singlet-triplet-singlet, strictly singlet, and strictly triplet He equiensembles.}
\label{fig:He:potxc:compare}
\end{figure}
Fig. \ref{fig:He:potxc:compare} compares $v\xcw(\br)$ for four types of He equiensembles, highlighting their different features.  The characteristic bump up in these potentials is shifted left in the 2-multiplet case, relative to the others shown.  This shift has little impact on the first ``shell" of the ensemble density's shell-like structure, but the second is shifted left and has sharper decay, noticeably different from that of the singlet ensemble.\cite{YTPB14}

\begin{table}
\begin{tabular}{cccc}
\hline\hline
\multicolumn{4}{l}{2-multiplet ensemble: $\omega_1=19.8231$ eV}\\
\hline
$\wt_2$ & 0.25 & 0.125 & 0.03125\\
$E_{1,\wt_2}^\text{KS}-E_{0,\wt_2}^\text{KS}$ & 25.1035 & 22.4676 & 21.6502\\
$\partial E_{\sss{xc},\wt_2}[n]/\partial \wt_2|_{n=n_{\wt_2}}$ & -15.8099 & -7.9358 & -5.4351\\
$(E_1-E_0)_{\wt_2}$ & 19.8336 & 19.8224 & 19.8385\\
\hline
\multicolumn{4}{l}{3-multiplet ensemble: $\omega_2=20.6191$ eV}\\
\hline
$\wt_3$ & 0.2 & 0.1 & 0.025\\
$E_{2,\wt_3}^\text{KS}-E_{0,\wt_3}^\text{KS}$ & 26.8457 & 25.8895 & 25.2853\\
$\partial E_{\sss{xc},\wt_3}[n]/\partial \wt_3|_{n=n_{\wt_3}}$ & -0.9596 & -0.7207 & -0.5696\\
$(E_2-E_0)_{\wt_2,\wt_3}$ & 20.6270 & 20.6184 & 20.6306\\
\hline
\multicolumn{4}{l}{triplet ensemble: $\omega_1=2.8991$ eV}\\
\hline
$\wt$ & 0.16667 & 0.08333 & 0.02083\\
$E_{1}^\text{KS}-E_{0}^\text{KS}$ & 2.8928 & 2.8956 & 2.8967\\
$\partial E\xcw[n]/\partial \wt|_{n=n_{\wt}}$ & 0.0187 & 0.0104 & 0.0074\\
$(E_1-E_0)_{\wt}$ & 2.8990 & 2.8990 & 2.8992\\
\hline\hline
\end{tabular}
\caption{He atom excitation energies, calculated using Eq. \parref{eqn:theory:exciteng} and various ensemble types: singlet-triplet (2-multiplet), singlet-triplet-singlet (3-multiplet), and strictly triplet. All energies are in eV. $\wt_2$ dependency of the 3-multiplet excitation energies is noted explicitly, though $\wt_2=(1-\wt_3)/4$ for the GOK ensemble.  See Supplemental Material for additional data and figures.}
\label{table:He:3state}
\end{table}

\begin{figure}[htbp]
\includegraphics[height=\columnwidth,angle=-90]{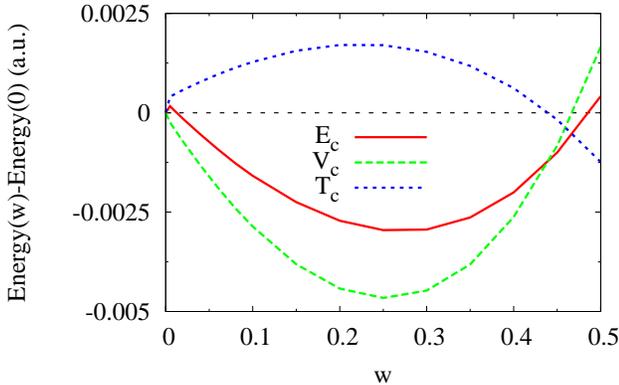}
\caption{Behaviors of the various energy components for the singlet ensemble of He. The ground state ($\wt=0$) values are taken from Ref. \onlinecite{HU97}. The small kinks near $\wt=0$ are due to the difference in the numerical approaches of this work and Ref. \onlinecite{HU97}.}
\label{fig:He:singlet:Ecomponents}
\end{figure}

The inequalities shown in Sec. \ref{sect:theory:inequ} and the virial theorem Eq. \parref{eqn:theory:virial:xc} are verified by the exact results. Behaviors of the energy components for the singlet ensemble versus $\wt$ are plotted in Fig. \ref{fig:He:singlet:Ecomponents}. Correlation energies show strong non-linear behavior in $\wt$. According to Eq. \parref{eqn:theory:exciteng}, the excitation energies are related to the derivative of $E\xc$ versus $\wt$. Therefore, $E\c$ is crucial for accurate excitation energies, even though its absolute magnitude is small.

\section{Conclusion}
\label{sect:conclusion}
This paper is an in-depth exploration of ensemble DFT, an alternative
to TDDFT for extracting excitations from DFT methodology.  Unlike
TDDFT, EDFT is based on a variational principle, and so one can
expect that the failures and successes of approximate functionals
should occur in different systems than those of TDDFT.

Apart from exploring the formalism and showing several new results,
the main result of this work is to apply a new algorithm to 
highly-accurate densities of eigenstates to explore the exact
EDFT XC potential. We find intriguing characteristic features of the exact potentials
that can be compared against the performance of old and new approximations.
We also extract the weight-dependence of the KS eigenvalues,
which are needed to extract accurate transition frequencies, and
find that a large cancellation of weight-dependence occurs in the
exact ensemble.  Many details of these calculations are reported in
the supplemental information.

From the original works of Gross, Oliviera, and Kohn,
ensemble DFT has been slowly developed over three decades by a few brave pioneering
groups, most prominently that of Nagy.  We hope that the insight these
exact results bring will lead to a plethora of new ensemble approximations
and calculations and, just possibly, a competitive method to
treating excitations within DFT.

\section*{Acknowledgements}
Z.-H.Y. and C.U. are funded by National Science Foundation Grant No. DMR-1005651. A.P.J. is supported by DOE grant DE-FG02-97ER25308. J.R.T. and R.J.N. acknowledge financial support from the Engineering and Physical Sciences Research Council (EPSRC) of the UK. K.B. is supported by DOE grant DE-FG02-08ER46496.

\nocite{Supp}
\bibliographystyle{unsrt}

\begin{thebibliography}{10}
\bibitem{note1}
There appears to be a sign error in Eq. (16) of Ref. \onlinecite{L95}: the two
  terms on the right-hand side should be swapped.
\end{thebibliography}

\end{document}